\documentclass[%
 aip,
 amsmath,amssymb,
reprint,
twocolumn,%
author-numerical,%
]{revtex4-1}
\usepackage{graphicx}
\usepackage{dcolumn}
\usepackage{bm}
\usepackage[utf8]{inputenc}
\usepackage[T1]{fontenc}
\usepackage{mathptmx}
\usepackage{amsmath}
\usepackage{amssymb}
\usepackage{amsfonts}
\usepackage{xcolor}

\usepackage[all]{background} 
\usepackage{url}
\SetBgContents{Accepted by J. Renew. Sustain. Energy}
\SetBgColor{gray}
\SetBgPosition{2.4,-7.0}
\SetBgOpacity{0.75}
\SetBgAngle{0.0}
\SetBgScale{3.5}
\usepackage{hyperref} 

\begin{document}

\preprint{AIP/123-QED}

\title[Effect of low-level jet height on wind farm performance]{\color{black}Effect of low-level jet height on wind farm performance}

\author{Srinidhi N. Gadde}
 \email[]{s.nagaradagadde@utwente.nl}
\author{Richard J. A. M. Stevens}%
  \email[]{r.j.a.m.stevens@utwente.nl}
\affiliation{Physics of Fluids Group, Max Planck Center Twente for Complex Fluid Dynamics, J. M. Burgers Center for Fluid Dynamics and MESA+ Research Institute, University of Twente, P. O. Box 217, 7500 AE Enschede, The Netherlands}

\date{\today}

\begin{abstract}
Low-level jets (LLJs) are the wind maxima in the lowest 50 to 1000 m of atmospheric boundary layers. Due to their significant influence on the power production of wind farms it is crucial to understand the interaction between LLJs and wind farms. In the presence of a LLJ, there are positive and negative shear regions in the velocity profile. The positive shear regions of LLJs are continuously turbulent, while the negative shear regions have limited turbulence. We present large eddy simulations of wind farms in which the LLJ is above, below, or in the middle of the turbine rotor swept area. We find that the wakes recover relatively fast when the LLJ is above the turbines. This is due to the high turbulence below the LLJ and the downward vertical entrainment created by the momentum deficit due to the wind farm power production. This harvests the jet's energy and aids wake recovery. However, when the LLJ is below the turbine rotor swept area, the wake recovery is very slow due to the low atmospheric turbulence above the LLJ. The energy budget analysis reveals that the entrainment fluxes are maximum and minimum when the LLJ is above and in the middle of the turbine rotor swept area, respectively. Surprisingly, we find that the negative shear creates a significant entrainment flux upward when the LLJ is below the turbine rotor swept area. This facilitates energy extraction from the jet, which is beneficial for the performance of downwind turbines.
\end{abstract}
\keywords{Low-level jet; Wind farm; Large eddy simulation; Stable boundary layer; Energy entrainment}

\maketitle

\section{Introduction}\label{sec1}
\begin{figure*}
 \centering
 \includegraphics[width=0.9\linewidth]{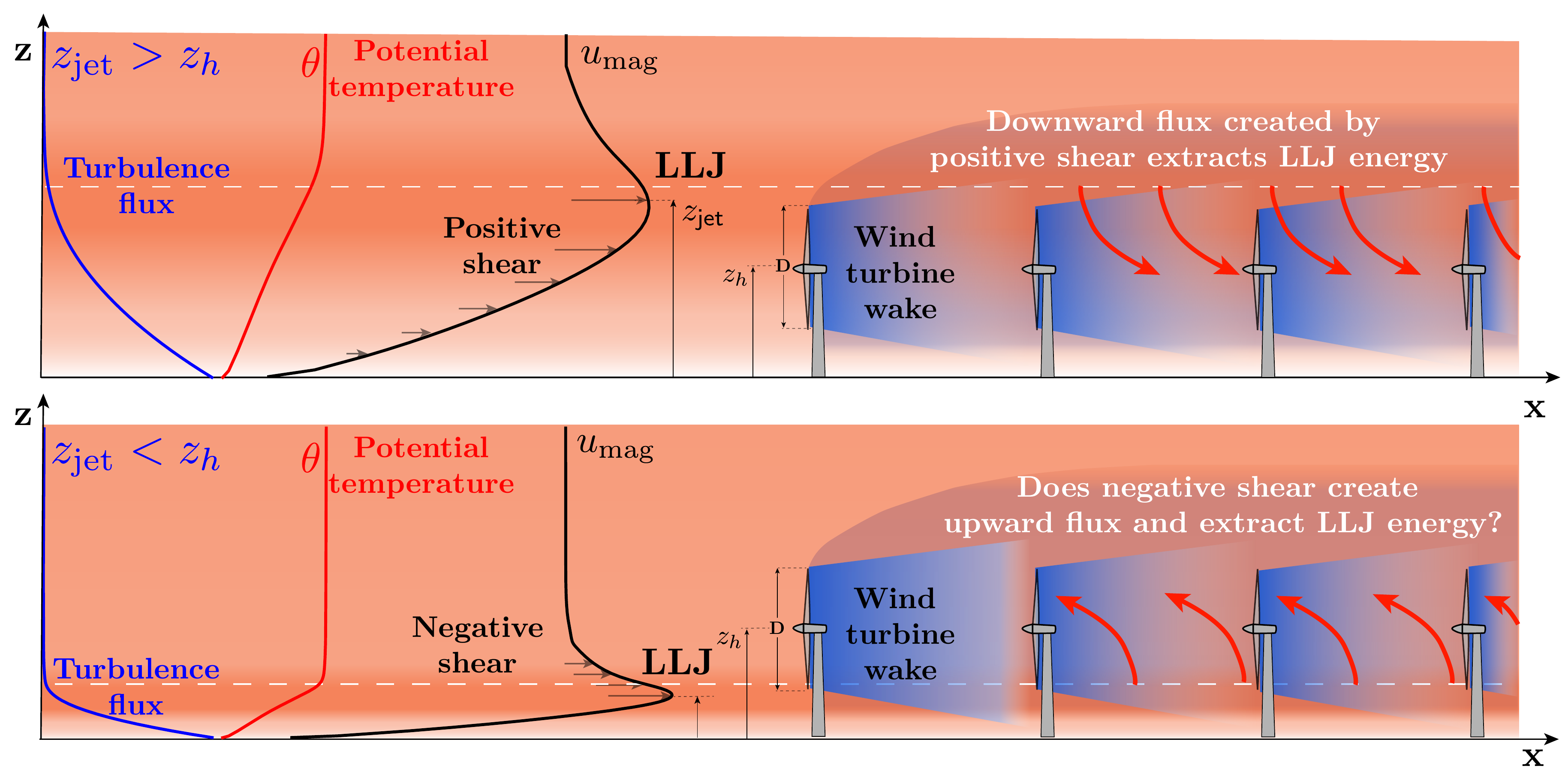}
 \caption{{\color{black} Problem definition. The top figure shows turbines operating in positive shear when the LLJ is above the turbine rotor swept area ($z_\mathsf{jet} > z_h$). The bottom figure shows turbines operating in negative shear when the LLJ is below the turbine rotor swept area ($z_\mathsf{jet} < z_h$). We study whether the negative shear increases the energy extraction from the jet by creating an upward flux. The figure also shows that the turbulent momentum flux is negligible above the LLJ. The potential temperature $\theta$ profile reveals that the boundary layer is stably stratified and shows a residual layer with a constant temperature above the LLJ.}}
 \label{fig1}
\end{figure*}

{\color{black} A low-level jet (LLJ) is the maximum in the wind velocity profile in the atmospheric boundary layer (ABL). {\color{black}When the wind in the residual layer\cite{bru82} is decoupled from the surface friction and subjected to inertial oscillations, the flow in the residual layer accelerates to super-geostrophic magnitudes and forms a LLJ \cite{bla57}.} These jets are observed in the lowest 50 to 1000 m of the ABL \citep{sme96} and are most pronounced in weak to moderately stable ABLs \cite{baa09, ban08}.} Figure \ref{fig1} shows a sketch of the velocity, potential temperature, and turbulence flux profiles in a stable ABL. LLJs \cite{kel04, ban02, sme93} have been reported all over the world with frequent occurrences in India \cite{pra11}, China \cite{liu14}, the Great Plains of the United States \cite{arr97} and the North Sea region of Europe \cite{kal19, wag19}. Field observations show that in the IJmuiden region of North Sea, LLJs are observed with a frequency of 7.56\% in summer and 6.61\% in spring\cite{dun18, kal17}. {\color{black} LLJs in the North Sea are associated with shallow boundary-layer heights \cite{dun18, baa09}, i.e.\ these jets can influence the wind farm power production. It has been reported that LLJs can increase the capacity factors by 60\% under nocturnal conditions \cite{wil15b}, and measurements in Western Oklahoma\cite{gre09} indicate that LLJs increase the power production compared to the case without jets.} As a result, the importance, relevance, and urgency of research into LLJ for wind farm applications have been outlined by van Kuik et al.\ \cite{kui16} in their long term European Research Agenda and a recent review by Port\'e-Agel et al.\ \cite{por20}.\\
\indent It is well established in the wind energy community that LLJs affect the performance of wind turbines \cite{sis78}. Below the jet height ($z_\text{jet}$) the velocity profile has a positive shear, and above the jet height there is a negative shear. The top panel of Fig.\ \ref{fig1} shows a turbine with hub-height ($z_h$) lower than the jet height, i.e.\ $z_\text{jet} > z_h$, operating in the positive shear region. The bottom panel shows a turbine with the hub-height higher than the jet height, i.e.\ $z_\text{jet} < z_h$, operating in the negative shear region. The potential temperature profile shows significant surface inversion with a residual layer above. Above the surface inversion the boundary layer has negligible turbulence and the region is associated with the negative shear, see Fig. \ref{fig1}.\\
\indent As noted above, LLJs generally form at the top of stable surface inversions \cite{baa09}, above which the turbulence is negligible \cite{bla57}. {\color{black}During an LLJ event, the turbulence intensity and turbulence kinetic energy are lower than for unstable conditions \cite{gut16}.} The effect of LLJs on wind turbine and wind farm performance has been studied before. Lu \& Port{\'e}-Agel \cite{lu11} performed large eddy simulations (LES) of an `infinite' wind farm in the stable boundary layer, and they report the formation of non-axisymmetric wakes and a decrease in the LLJ strength due to the energy extraction by the turbines. {\color{black}LLJ elimination due to wind turbine momentum extraction has also been reported in similar LES studies \cite{abk16, bha15, sha17d, ali17}.} Also, mesoscale simulations in which wind farms are modeled as localized roughness elements show that LLJs are eliminated by wind farms\cite{fit13}. Furthermore, due to the velocity maximum and strong shear, both the power production and the fatigue loads on wind turbines are affected by the LLJ \cite{gut17}.\\
\indent In a fully developed wind farm boundary layer, the power production depends on the vertical entrainment fluxes from above, which is created by the momentum deficit inside the wind farm. Wind turbines operating below the LLJ are subjected to positive shear and continuous turbulence. In that case, the downward entrainment fluxes are enhanced due to which the energy of the LLJ is harvested \cite{lu11, na18, nag20b}. Recently, \citeauthor{doo20}\cite{doo20} conducted experiments in which they studied the interaction between a wind farm and a synthetic jet, they report that entrainment fluxes are enhanced due to the shear layer associated with the LLJ. However, situations might arise when the turbines have to operate in the negative shear region with reduced turbulence. Aeroelastic simulations of the interaction between a LLJ and a wind turbine show that the loading on a wind turbine decreases when it operates in the jet's negative shear region. Based on these simulations, Gutierrez et al.\ \cite{gut16} suggest installing turbines at heights where negative shear occurs. However, the region where negative shear occurs is also a region of reduced turbulence{\color{black}, which negatively affects wake recovery and hence the power production of downwind turbines. \cite{ste17, por20, ali18b}}\\
\indent {\color{black} Previous studies have mostly focused on wind turbines and wind farms operating in the jet's positive shear region. The presence of negative shear and negligible turbulence above the LLJ leads to scenarios in which the wind farm and LLJ interaction is not straightforward. As mentioned before, when the LLJ is above the turbines, the momentum deficit creates a downward entrainment flux, which facilitates the extraction of LLJ's energy, see the top schematic in Fig. \ref{fig1}. However, it is not yet explored if the negative shear creates an upward entrainment flux when the LLJ is below the turbine rotor swept area, see the bottom schematic in Fig. \ref{fig1}. Therefore, the objective of the study is to understand how changing the LLJ height relative to the turbine hub-height affects wake recovery and power production of downstream turbines. This can be done in two ways: by keeping the jet height constant and changing the turbine height or simulating jets of different heights which involves multiple spin-up simulations with different boundary layer properties. Changing jet height is complicated and computationally expensive, therefore we follow the easiest and the most straightforward way of changing the ratio $z_\text{jet} < z_h$ by changing the turbine height. In this work, we consider three different scenarios:
\begin{itemize}
\item[1.] the LLJ above the turbine rotor swept area, i.e.\ $z_\text{jet} < z_h$.
\item[2.] the LLJ in the middle of the turbine rotor swept area, i.e.\ $z_\text{jet} \approx z_h$.
\item[3.] the LLJ below the turbine rotor swept area, i.e.\ $z_\text{jet} > z_h$.
\end{itemize}
}
In section \ref{sec2}, the simulation methodology and the wind farm configuration are described. In section \ref{sec3} the main observations are discussed and in section \ref{sec4} the major conclusions are detailed. 
\section{Simulation methodology}\label{sec2}

\subsection{Governing equations}\label{sec2.1}
We numerically integrate the filtered Navier-Stokes equation coupled with the Boussinesq approximation to model buoyancy. The governing equations are: 
\begin{align}
 \partial_{\mathit{i}} \widetilde{u}_\mathit{i}&=0,\label{eqn1}\\
 \begin{split}
 \partial_{\mathit{t}}\widetilde{u}_\mathit{i} + \partial_\mathit{j}\left(\widetilde{u}_\mathit{i}\widetilde{u}_\mathit{j}\right)&=-\partial_{\mathit{i}}\widetilde{p}-\partial_\mathit{j}\tau_{\mathit{ij}} + g\beta(\widetilde{\theta}-\widetilde{\theta}_\mathit{0})\delta_{\mathit{i3}}\\&+f_c(U_g-\widetilde{u})\delta_{i2}-f_c(V_g-\widetilde{v})\delta_{i1}+ \widetilde{f}_x\delta_{i1}+ \widetilde{f}_y\delta_{i2},
 \end{split}\label{eqn2}\\
 \partial_{\mathit{t}}\widetilde{\theta} + \widetilde{u}_\mathit{j}\partial_{\mathit{j}}\widetilde{\theta}&=-\partial_{\mathit{j}}q_\mathit{j},\label{eqn3}
\end{align}
here the tilde represents a spectral cut-off filter of size $\Delta$, $\widetilde{u}_\mathit{i}=\left(\widetilde{u},\widetilde{v},\widetilde{w}\right)$ and $\widetilde{\theta}$ are the filtered velocity and potential temperature, respectively, $g$ is the gravitational acceleration, $\beta=1/\theta_\mathit{0}$ is the buoyancy parameter with respect to the reference potential temperature $\theta_\mathit{0}$, $\delta_{\mathit{ij}}$ is the Kronecker delta, and $f_c$ is the Coriolis parameter. The ABL is forced by a mean pressure $p_\infty$, represented by the geostrophic wind with the relation, $U_g=-\frac{1}{{\rho}f_c}\frac{\partial{p_{\infty}}}{\partial{y}}$ and $V_g=\frac{1}{{\rho}f_c}\frac{\partial{p_{\infty}}}{\partial{x}}$ as its components. $\widetilde{p}=\widetilde{p}^{*}/\rho+\sigma_{kk}/3$ is the modified pressure, which is the sum of the trace of the SGS stress, $\sigma_{kk}/3$, and the kinematic pressure $\widetilde{p}^{*}/\rho$, where $\rho$ is the density of the fluid. 

It is well established that the actuator disk model can capture the wake dynamics starting from $1$ to $2$ diameters downstream of the turbine sufficiently accurately \cite{ste17, ste18, wu11}. Therefore, the actuator disk model can be used to study the large scale flow phenomena in a wind farm on which we focus here. We note that the actuator disk model cannot capture the vortex structures near the turbine due to the absence of the turbine blades\cite{sor11, tro10, ste17}. To capture vortex structures very high resolution actuator line model simulations are required, which is not feasible for large wind farms \cite{ste18}. Therefore, we use a well-validated actuator disk model \cite{jim07, jim08,cal10, ste14, ste16, zha19,nag19} in this study. The turbine forces $\widetilde{f}_x$ and $\widetilde{f}_y$ in equation \eqref{eqn2} are modeled using the turbine force
\begin{equation}
 F_t = -\frac{1}{2}\rho{C_T}{U^2_\infty}\frac{\pi}{4}D^2,\label{eqn:force}
\end{equation} 
where $C_T$ is the thrust coefficient and $U_\infty$ is the upstream undisturbed reference velocity. Equation \eqref{eqn:force} is only applicable for isolated turbines \cite{jim07, jim08}. In wind farm simulations the upstream velocity $U_\infty$ cannot be readily specified. Consequently, it is common practice \cite{cal10, cal11} to use actuator disk theory to relate $U_\infty$ with the rotor disk velocity $U_d$,
\begin{equation}
U_\infty= \frac{U_d}{\left(1-a\right)}\label{eqn:uinfty}
\end{equation} 
where $a$ is the axial induction factor. The turbine forces are calculated by substituting equation \eqref{eqn:uinfty} in equation \eqref{eqn:force}. For a detailed description and validation of the employed actuator disk model we refer the reader to Refs.\ \cite{cal10, cal11,ste18}.

The terms involving molecular viscosity are neglected due to the high Reynolds number of the ABL flow. $\tau_{\mathit{ij}}=\widetilde{u_{\mathit{i}}u_{\mathit{j}}}-\widetilde{u}_\mathit{i}\widetilde{u}_\mathit{j}$ is the traceless part of the SGS stress tensor and $q_\mathit{j}=\widetilde{u_\mathit{j}\theta}-\widetilde{u}_\mathit{j}\widetilde{\theta}$ is the SGS heat flux tensor. The SGS stresses and heat fluxes are modeled as,
\begin{align}
 \tau_{\mathit{ij}}&=\widetilde{u_{\mathit{i}}u_{\mathit{j}}}-\widetilde{u}_\mathit{i}\widetilde{u}_\mathit{j}=-2\nu_{T}\widetilde{S}_{ij}=-2(C_s\Delta)^2|\widetilde{S}|\widetilde{S}_{ij},\label{eqn4}\\
 q_\mathit{j}&=\widetilde{u_\mathit{j}\theta}~~-\widetilde{u}_\mathit{j}\widetilde{\theta}~~=-\nu_\theta\partial_j\widetilde{\theta}~~=-(D_s\Delta)^2|\widetilde{S}|\partial_j\widetilde{\theta},\label{eqn5}
\end{align}
where $\widetilde{S}_{ij}=\frac{1}{2}\left(\partial_j{\widetilde{u}_i} + \partial_i{\widetilde{u}_j}\right)$ is the grid-scale strain rate tensor, $\nu_T$ is the eddy viscosity, $C_s$ is the Smagorinsky coefficient for the SGS stresses, $\Delta$ is the grid size, $\nu_\theta$ is the eddy heat diffusivity, $D_s$ is the Smagorinsky coefficient for the SGS heat flux, and $|\widetilde{S}| = \sqrt{2\widetilde{S}_{ij}\widetilde{S}_{ij}}$. To model the SGS stresses without any ad-hoc modifications, we use a tuning-free, scale-dependent, dynamic model based on the Lagrangian averaging of the coefficients \cite{bou05, sto06, sto08}. The model has been found to be highly suitable for inhomogeneous flows such as the flow inside wind farms\cite{ste16}. For further details and validation of the code we refer to Gadde et al.\ (2020) \cite{nag20}. 

\subsection{Numerical method}\label{sec2.2} 
We use a standard pseudo-spectral method to calculate the derivatives in horizontal directions and a second-order central difference scheme to calculate the gradients in the vertical direction. A second-order Adams-Bashforth scheme is employed to advance the solution in time. The aliasing errors in the non-linear terms are removed by the 3/2 anti-aliasing method\cite{can88}. The advective terms in the governing equations are written in the rotational form\cite{fer02}. We discretize the horizontal directions uniformly with $n_x$, $n_y$, grid points in the streamwise and spanwise directions, respectively. This results in grid sizes of $\Delta_{x}=L_x/n_x$, $\Delta_{y}=L_y/n_y$ in the horizontal directions. {\color{black} In the vertical direction, we use a uniform grid up to a certain height, above which we use a stretched grid. The vertical grid size in the uniform region of the computational domain is represented by $\Delta_z$.} The horizontal and vertical computational planes are staggered, such that for the horizontal velocity components, the first vertical grid point above the ground is located at $\Delta_{z}/2$. No-slip and free-slip boundary conditions are imposed at the lowest and the topmost computational plane, respectively. We use the Monin-Obukhov similarity theory \cite{moe84} to model the instantaneous stress and heat flux at the wall by using the velocity and temperature at the first grid point above the wall
\begin{align}
\tau_{i3|w}=-{u_{*}^2}\frac{\widetilde{u}_i}{\widetilde{u}_r}=-\Bigg(\frac{\widetilde{u}_r\kappa}{\text{ln}(\Delta{z}/2z_o)-\psi_{M}}\Bigg)^2\frac{\widetilde{u}_i}{\widetilde{u}_r},\label{eqn6}
\end{align}
and
\begin{align} 
q_{*}&=\frac{u_{*}\kappa(\theta_s-\widetilde{\theta})}{\text{ln}(\Delta{z}/2z_{os})-\psi_{H}}.\label{eqn7}
\end{align}
\noindent In the above equations, $\widetilde{u}_i$ and $\widetilde{\theta}$ represent the filtered grid-scale velocities and potential temperature at the first grid point above the ground, $u_*$ is the frictional velocity, $z_o$ is the roughness height for momentum, {\color{black} $z_{os}$ is the roughness height for heat flux}, $\kappa=0.4$ is the von K\'arm\'an constant, $\widetilde{u}_r=\sqrt{\widetilde{u}^2 + \widetilde{v}^2}$ is the resolved velocity magnitude, and $\theta_s$ is the grid scale potential temperature at the surface. $\psi_M$ and $\psi_H$ are the stability corrections for momentum and heat flux, respectively. We use the stability correction used by Beare et al.\ \cite{bea06} to simulate the stable boundary layer, i.e.\ $\psi_{M}= -4.8z/L$ and $\psi_{H}= -7.8z/L$, where $L=-({u_*}^3\theta_{0})/({\kappa}gq_{*})$ is the surface Obukhov length. We note that, for convenience, the tildes representing filtered LES quantities are omitted in the remainder of the paper.

\subsection{Boundary layer characteristics}\label{sec2.3} 
\begin{table*} 
 \begin{center}
 \caption{The table gives the size of the computational domain and the used grid resolution in the streamwise ($n_x$), spanwise ($n_y$), and vertical ($n_z$) direction, respectively. $C_r$ is the surface cooling rate, $z_i$ is the boundary layer height, $z_\text{jet}$ is the jet height, $u_*$ is the friction velocity, $u_\text{jet}/G$ is the non-dimensionalized jet velocity, and $z_i/L$ represents the stability parameter.}
 \begin{tabular}{|c|c|c|c|c|c|c|c|c|c|c|}
 \hline
 Domain size & $n_x \times n_y \times n_z$ & $C_r$ [$\text{K}\cdot\text{h}^{-1}$] & $z_i$ [m] & $z_\text{jet}$ [m] & $u_*$ [m$\text{s}^{-1}$]& $u_\text{jet}/G$ & $z_i/L$ \\[3pt]
 \hline
$11.52$ $\text{km}$ $\times$ $4.6$ $\text{km}$ $\times$ $3.84$ $\text{km}$ & $1280\times512\times384$& 0.50 & 131.6 & 125 & 0.192 & 1.21 & 2.95\\
\hline
 \end{tabular}
\label{table1}
\end{center}
\end{table*}

\begin{figure*}
 \centering
 \includegraphics[width=\linewidth]{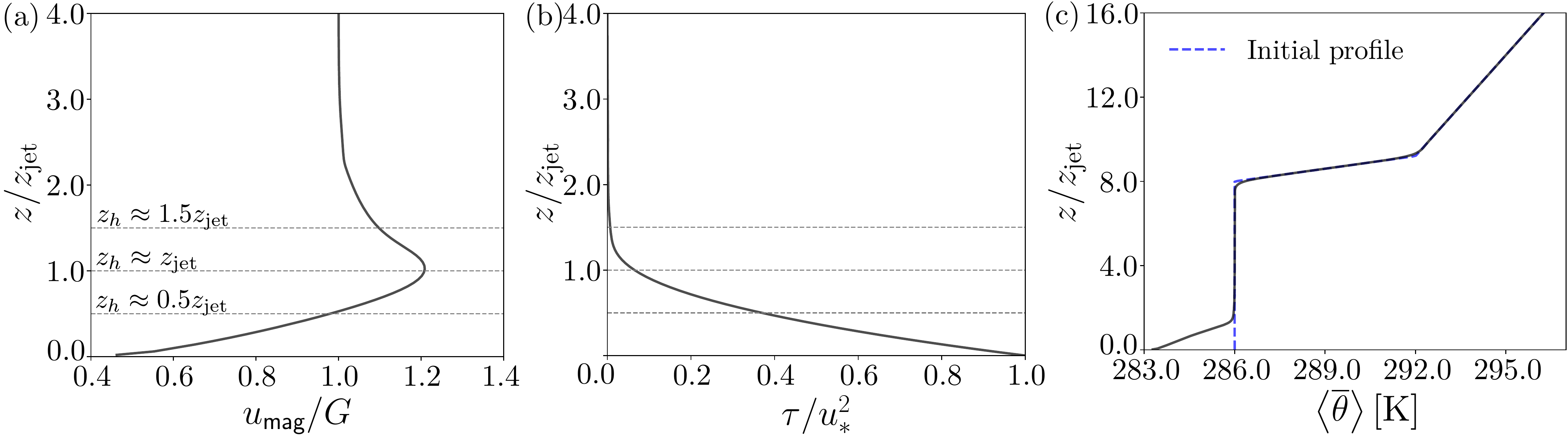}
 \caption{(a) Horizontally averaged wind magnitude $u_\text{mag}/G$, (b) the vertical momentum flux, and (c) the temperature profiles plotted as a function of the height. Height is normalized with the jet height.}
 \label{fig2}
\end{figure*}
\vspace{-5mm}
{\color{black} We consider a continuously turbulent, moderately stable ABL with a capping inversion at approximately 1000 m. The temperature profile is slightly modified form of the one used in the LES of second Global Earth and Water Cycle Experiment (GEWEX) ABL study (GABLS-2) single column intercomparison setup \cite{kum10}. The boundary layer is initialized with a constant temperature of 286 $\text{K}$ below 1000 $\text{m}$, a capping inversion of strength 6 $\text{K}$ between 1000 m and 1150 m, followed by a constant temperature gradient of 5 $\text{K}\cdot\text{km}^{-1}$ above. The initial temperature profile is shown by the blue dashed lines in Fig.\ \ref{fig2}(c). The roughness height is, $z_o=0.002$ m for momentum corresponding to offshore conditions \cite{dor15} and $z_{os} = \frac{z_o}{10}= 0.0002$ m \cite{bru82} for modelling the heat flux. The surface is cooled at a constant rate of 0.5 $\text{K}\cdot\text{hour}^{-1}$. The geostrophic forcing is set to $G=(U_g,V_g)=(8.0,0.0)$ $\text{m}\text{s}^{-1}$ and the Coriolis parameter is set to $f_c=1.159\times10^{-4}$ $\text{s}^{-1}$ corresponding to a latitude of $52.8^\circ$, which is representative for the Dutch North Sea. The velocity is initialized with the geostrophic velocity, and uniform random perturbations are added to the initial velocities and temperature up to a height of 500 m to trigger turbulence. We note that the boundary layer reaches a quasi-steady state at the end of 8$^\text{th}$ hour.}\\
\indent The boundary layer characteristics relevant to this study are given in table \ref{table1}. {\color{black} The jet height $z_\text{jet}$ is approximately $125$ m. The boundary layer height, defined as the height where the shear stress reaches 5\% of its surface value\cite{bea06}, is 131.6 m. The ratio of boundary layer height to the surface Obukhov length is $z_i/L=2.95$. Scaling regimes reported by \citeauthor{hol86} \cite{hol86} show that for $z_i/L < 3$, stable boundary layers show negligible intermittency throughout the boundary layer. This confirms that the boundary layer is moderately stable \cite{hol86}. The jet velocity is 9.68 m/s. It is worth mentioning here that LLJs with heights between 80 and 200 m and jet velocities of $8$ to $10$ m/s are frequently observed in the Dutch North Sea region \cite{baa09}.}\\
\indent Figure \ref{fig2}(a) shows the horizontally averaged velocity magnitude $u_\text{mag}=\left<\sqrt{\overline{u}^2+\overline{v}^2}\right>$ variation with height and Fig.\ \ref{fig2}(b) the corresponding horizontally averaged vertical turbulent momentum flux $\tau=\left<\sqrt{(\overline{u'w'})^2 + (\overline{v'w'})^2}\right>$, where $\overline{u'w'}=\left(\overline{{uw}} + \overline{\tau_{xz}}\right)-\overline{{u}}~\overline{{w}}$ and $\overline{v'w'}=\left(\overline{{vw}} + \overline{\tau_{yz}}\right)-\overline{{v}}~\overline{{w}}$. {\color{black} This figure reveals that there is negligible turbulence above the jet. Figure \ref{fig2}(c) presents the horizontally averaged potential temperature with surface inversion top at approximately 140 m. The inversion height is defined as the height at which the temperature gradient is highest. The inversion top acts as a lid separating the turbulent and non-turbulent regions of the boundary layer. The temperature profile shows a prominent residual layer above the LLJ.}
\subsection{Computational domain and wind farm layout}\label{sec2.4} 
\begin{figure*}[ht!]
	\centering
	\includegraphics[width=\linewidth]{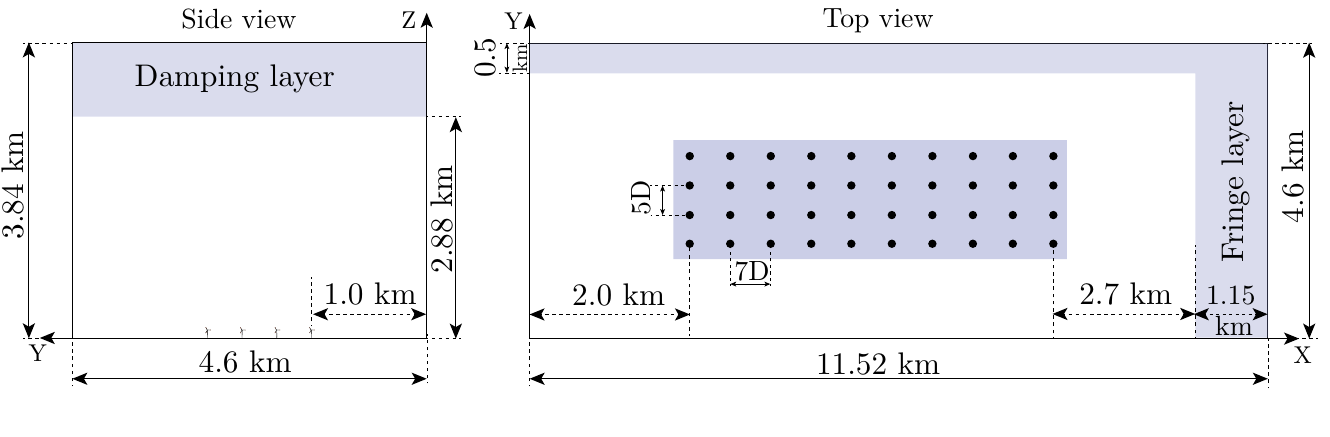}
	\caption{Schematic to show the wind farm layout with the damping layer to prevent gravity waves and the fringe layer used in the concurrent precursor method. The black circles denote the wind turbine locations. All the flow statistics are sampled from the shaded region with dimensions $70D\times{20D}\times{D}$ centered around the wind farm; see also figure \ref{cv}.}
	\label{fig3}
\end{figure*}

{\color{black} \indent The computational domain is $11.52$ $\text{km}$ $\times$ $4.6$ $\text{km}$ $\times$ $3.84$ $\text{km}$, which is discretized by $1280\times512\times384$ grid points. The grid points are uniformly distributed in horizontal directions. This leads to a uniform grid resolution of 9 m in both streamwise and spanwise directions. In the vertical direction a grid spacing of 5 m is used up to 1500 m, above which the grid is slowly stretched. Here we emphasize that our simulations benefit from using an advanced Lagrangian dynamic SGS scale model, which has been shown to capture the dynamics of stable boundary layers very well \cite{sto08, nag19}. It is worth mentioning here that in the LES intercomparison of the most widely studied stable boundary layer, Beare et al.\ \cite{bea06} report that a grid size of 6.25 m produces reasonably acceptable results compared to high-resolution LES of stable boundary layers. To provide perspective, recently, Allaerts and Meyers \cite{all18} in the simulation of wind farms in a stable boundary layer, used a horizontal resolution of 12.5 m and a vertical resolution of 5 m. Furthermore, Ali et al.\ \cite{ali17} in their simulations of wind farms in diurnal cycles, which also includes stable boundary layers, use a horizontal resolution of 24.5 m and a vertical resolution of 7.8 m. So the resolution employed here is relatively high for simulations of such large wind farms.}\\
\indent {We employ the concurrent precursor technique \cite{ste14} to introduce the inflow conditions sampled from the precursor simulation into the wind farm domain. LLJs generally occur over small regions and have limited spanwise width. Therefore, we use fringe layers in both streamwise and spanwise directions to remove the effect of periodicity. A Rayleigh damping layer\cite{kle78} with a damping constant of 0.016 $\text{s}^{-1}$ is used in the top 25$\%$ of the domain to damp out the gravity waves triggered by the wind farm.}\\
{\color{black} \indent We consider a wind farm with 40 turbines distributed in 4 columns and 10 rows, see Fig.\ \ref{fig2}. We choose turbines of diameter, $D=80$ m and the turbines are separated by a distance of $7D$ and $5D$ in the streamwise and spanwise directions, respectively. The objective of our study is to study the effect of $z_\text{jet}/z_h$ on wake recovery and wind farm power production. To achieve that, we vary the turbine hub height such that $z_\text{jet}>z_h$, $z_\text{jet}{\approx}z_h$, and $z_{\text{jet}} < z_h$, corresponding hub-heights are $z_h=0.5z_\text{jet}$, $z_\text{jet}$, $1.5z_\text{jet}$. The three cases represent the scenarios when the LLJ is below, above, and in the middle of the turbine rotor swept area. We perform two additional simulations with $z_h/D=0.75$ and turbine diameters 160 m and 240 m to study the effect of turbine diameter on wind farm performance. In these simulations, the turbines are separated by $720$ m in the streamwise direction. In the spanwise direction, for the cases with turbine diameter $160$ m and $240$ m, the turbines are separated by $480$ m and $720$ m, respectively.}\\
\indent Calaf et al.\ \cite{cal10} and Meyers and Meneveau \cite{mey13} showed that a resolution of $25$ m $\leq \Delta{x} \leq$ $50$ m in the streamwise direction and of $10$ m $\leq \Delta{y}\leq$ $25$ m in the spanwise direction is sufficient when an actuator disk method is used to model the turbines. Furthermore, Wu and Port\'e-Agel \cite{wu11} showed that one needs $8$ points along the diameter in the vertical direction and $5$ points along the diameter in the spanwise direction. Clearly, our simulations satisfy these criteria as we use a $5$ m resolution in the vertical and a $9$ m resolution in the horizontal directions. This means the turbine disk is discretized by $16$ points in the vertical and \ $9$ points in the spanwise direction, respectively.\\
\indent We use a proportional-integral (PI) controller\cite{all15} to ensure that the mean wind direction at hub-height is from West to East. The wind angle controller has been successfully used in our previous study of wind farms in neutral and stable boundary layers \cite{nag19} and ensures that the wind farm geometry is the same for all considered cases. Yaw misalignment due to the local changes in the wind angle is prevented by rotating the actuator disks such that the disks are always perpendicular to the local wind angle. Figure \ref{fig3} shows the wind farm layout and the dimensions of the different regions in the computational domain.

\section{Results \& discussions}\label{sec3}
The simulations were carried out in two stages. In the first stage, only the boundary layer in the precursor domain is simulated. After the quasi-steady conditions are reached, the turbines are introduced at the end of the 8$^\text{th}$ hour. In this second stage, the simulations are continued for two more hours, and the statistics are collected in the last hour. {\color{black} Each simulation costs about 0.3 million CPU hours.}  In section \ref{sec3.1} the flow structures are analyzed, followed by a discussion on power production, momentum flux, and wake recovery in section \ref{sec3.2}. In section \ref{sec3.3}, an energy budget analysis is presented in which we discuss the diverse processes affecting the wind farm performance in the presence of a LLJ.

\subsection{Flow structures}\label{sec3.1}
\begin{figure*}
 \centering
 \includegraphics[width=1.0\linewidth]{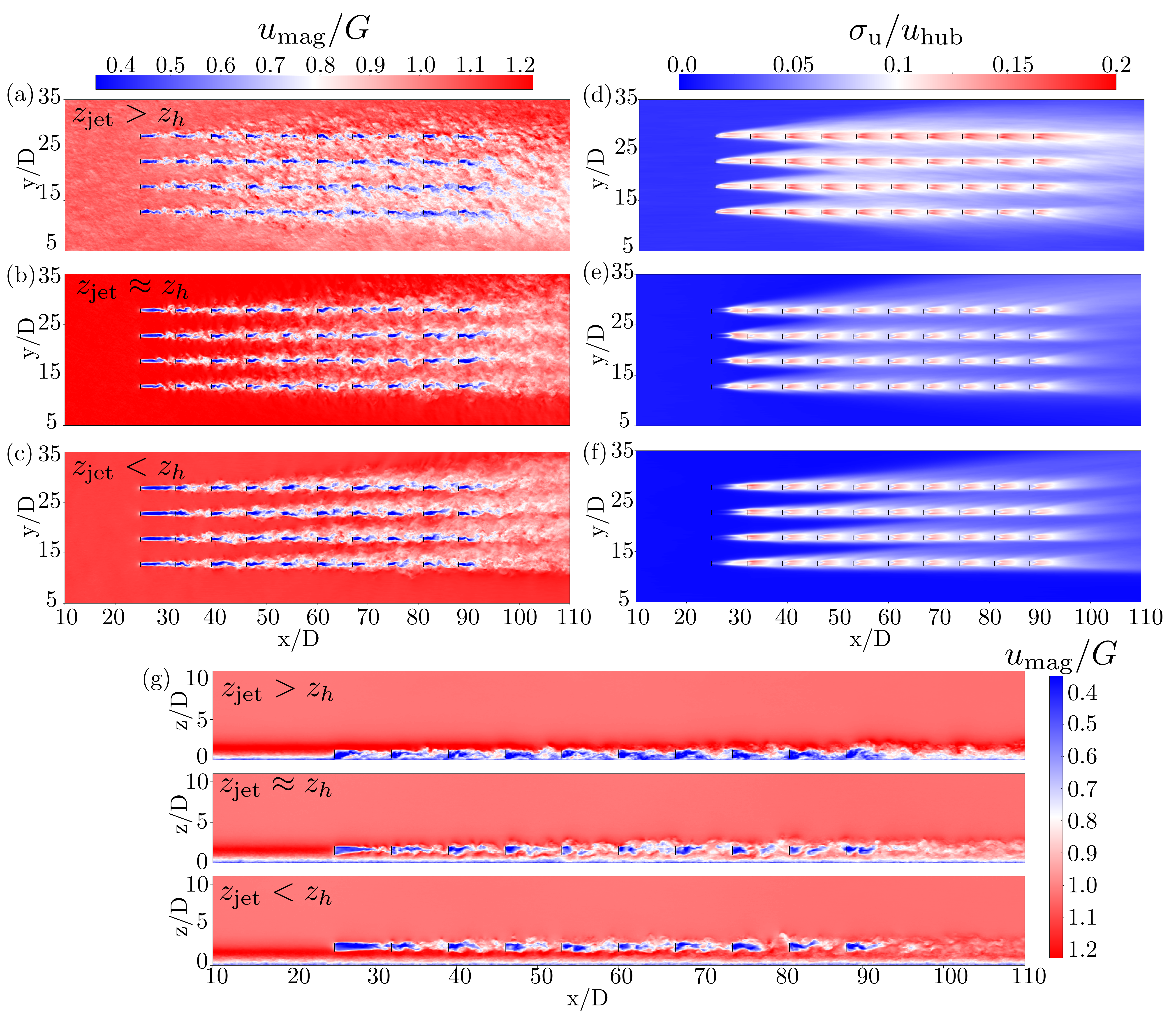}
 \caption{Normalized instantaneous velocity $u_\text{mag}/G$ at hub-height for the three cases, i.e.\ (a) $z_\text{jet} > z_h$, (b) $z_\text{jet}\approx z_h$, and (c) $z_\text{jet} < z_h$. Figs. (d), (e), and (f) present the corresponding time-averaged turbulence intensity, $\sigma_{u}/u_\text{hub}$ where $\sigma_{u}=\sqrt{2k/3}$ and k is the turbulent kinetic energy. (g) Side view of the instantaneous streamwise velocity in an x-z plane through the second turbine column for the different cases.}
 \label{topview}
\end{figure*}
\indent A visualization of the instantaneous velocity at hub-height is presented in Figs. \ref{topview}(a), (b), and (c). Figures \ref{topview}(d), (e), and (f) show the time-averaged turbulence intensity for all three cases. The turbulence intensity is calculated as $\sigma_u=\sqrt{2k/3}$, where $k=0.5 (\overline{u'^2} +\overline{v'^2} + \overline{w'^2})$ is the resolved turbulent kinetic energy and $u_\text{hub} = \sqrt{\overline{{u}}^2 + \overline{{v}}^2 + \overline{{w}}^2}$ {\color{black} is the velocity at the hub-height at the inlet.} {\color{black} When the LLJ is above the turbines,} small scale structures are visible in the entrance region in front of the wind farm, see Fig.\ \ref{topview}(a). In this case, the turbines operate in a completely turbulent region, and the wakes show significant turbulence towards the end of the wind farm, see Fig.\ \ref{topview}(d). The wakes recover relatively fast due to the high atmospheric turbulence and the additional wake generated turbulence. {\color{black} In contrast, Figs.\ \ref{topview}(b) and \ref{topview}(e) show less turbulence in the entrance region of the wind farm and behind the first turbine row for the $z_h \approx z_\text{jet}$ case.} {\color{black} However, towards the rear of the wind farm, we observe significant turbulence. {\color{black} This effect is prominent when the LLJ is below the turbines ($z_\text{jet} < z_h$), and we observe only marginal {\color{black} wake turbulence} behind the first turbine row, see Fig.\ \ref{topview}(c) and \ref{topview}(f). This will affect the wake recovery and consequently the power production of the second turbine row. The limited turbulence at hub-height at the farm entrance is due to the strong thermal stratification associated with the surface inversion top.} However, after the first couple of rows, we observe significant turbulence created by the wakes. It is widely accepted that the turbine wake meandering and corresponding wake turbulence is related to the atmospheric turbulence \cite{mao18, lar08}, and in the absence of atmospheric turbulence, the wake turbulence is also limited. In essence, the wake recovery is affected when turbines operate in the negative shear region above the LLJ.}\\ 
\begin{figure*}[ht!]
	\centering
	\includegraphics[width=0.9\linewidth]{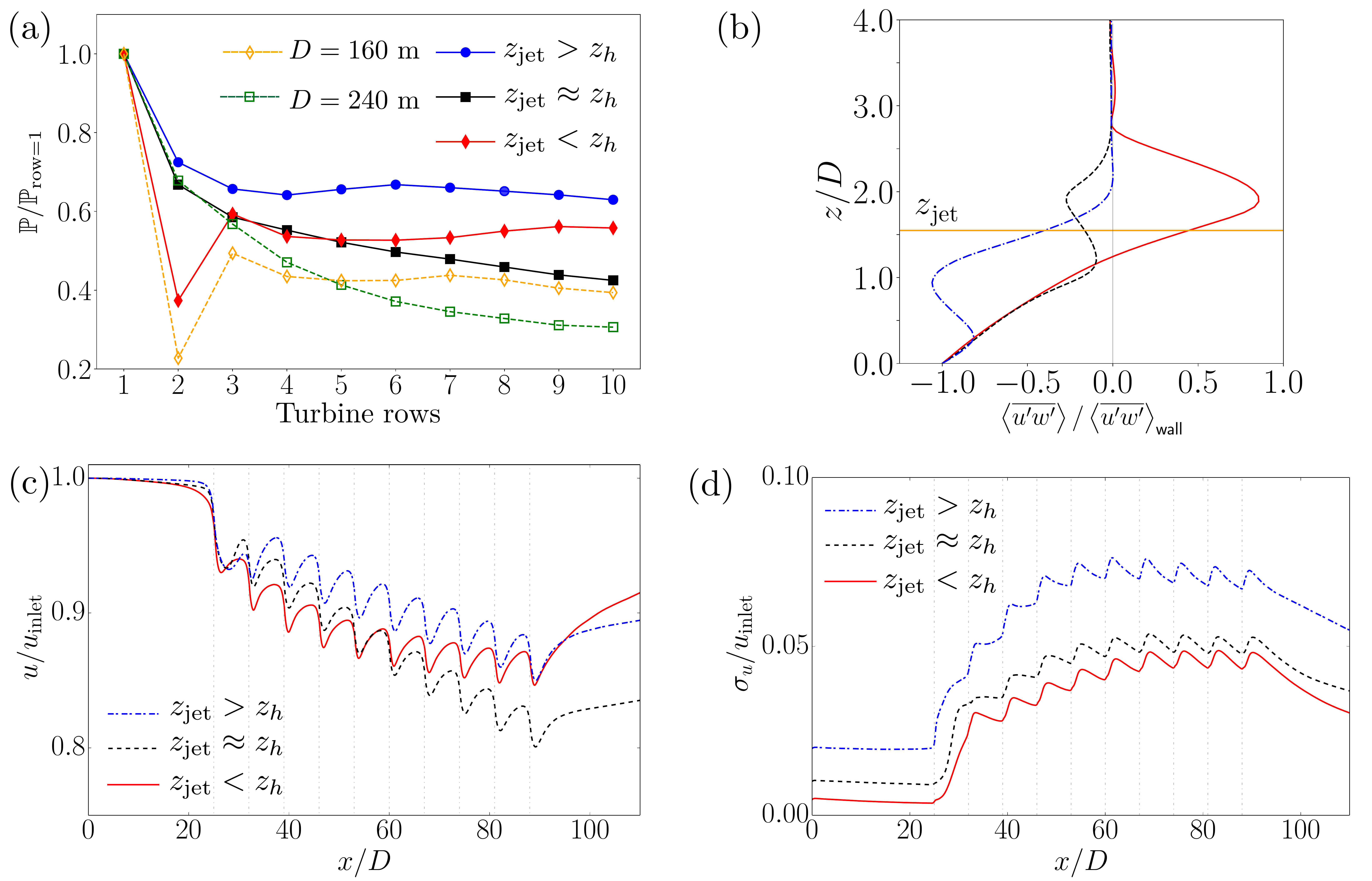} 
	\caption{(a) The row-averaged power normalized with the power production of the first row. Results from the additional simulations with $D=160$ m and $D=240$ m are also included. (b) Planar averaged streamwise vertical momentum flux versus height. (c) Spanwise averaged streamwise velocity normalized with the upstream velocity at hub-height as a function of the streamwise location. (d) Streamwise variation of the turbulence intensity at hub-height for the three different cases.}
	\label{powerplots}
\end{figure*}
\indent Figure \ref{topview}(g) shows the side view of the wind farm in an x-z plane passing through the second turbine column. The top panel in Fig.\ \ref{topview}(g) shows the turbines operating in a turbulent region. 
It is worth noting that the turbine wakes show significant wake turbulence, which aids the extraction of momentum from the LLJ. When $z_\text{jet} \approx z_h$, the turbines in the first row extract the energy in the jet, and turbines operate in a well-mixed region after the second turbine row. {{\color{black} Moreover, the LLJ reduces in strength after the first turbine row, and therefore rows that are further downstream cannot benefit from the jet anymore.} However, when the LLJ is below the turbines ($z_\text{jet} < z_h$), the first couple of turbine rows are in a non-turbulent region, and therefore the turbulence after the first turbine row is limited. Furthermore, we observe no transverse wake meandering behind the first turbine row due to low atmospheric turbulence in the thermally stratified region above the jet. The jet's strength is reduced due to energy extraction towards the rear of the wind farm because of the positive entrainment flux created from below due to the energy extraction by turbines. This will be discussed in detail in the next section.

\subsection{Power production and wake recovery}\label{sec3.2}

The row-averaged power normalized by the first row's power production is presented in Fig.\ \ref{powerplots}(a). The turbine power is averaged in the $10^\text{th}$ hour of the simulation. {\color{black} Results from the additional simulations with the diameters $160$ m and $240$ m are also included in the figure.}
{\color{black} When the LLJ is above the turbines ($z_\text{jet} > z_h$)}, we observe that the relative power production is higher, which means that velocity recovers faster than in the other cases, see the plot of wake recovery in Fig.\ 5(c). When $z_h \approx z_\text{jet}$, the power production continuously reduces towards the rear of the wind farm, see Fig.\ \ref{powerplots}(a). The corresponding wake recovery shows that the velocity continuously drops in the downstream direction, which indicates that the wake recovery is negligible; see the dashed line in Fig.\ \ref{powerplots}(c). Interestingly, when the jet is below the turbines ($z_\text{jet} < z_h$), the power production of the second row is severely affected due to the {\color{black} absence of turbulence in the wake of the first turbine row. However, the power production increases further downstream due to wake generated turbulence, and it shows an upward trend towards the back of the wind farm.} For this case, {\color{black}the wake turbulence becomes significant for $x/D > 30$ behind the second turbine row,} and subsequently, the turbines entrain high momentum wind from the LLJ, and the wake recovers significantly. {\color{black} For the additional cases with turbines with bigger diameter of 160 m and 240 m, we find that the overall trends in the normalized power production as a function of the downstream position remains the same even though the streamwise turbine spacing is small. This confirms that the results presented in this study capture the relevant physics of the different scenarios, i.e.\ when the  LLJ is below, in the middle, or above the turbine rotor swept area.}\\
\indent To understand the wake recovery and the associated power production of downstream turbines, the planar averaged vertical turbulent flux of streamwise momentum $\left<\overline{u'w'}\right>$ and normalized by the $\left<\overline{u'w'}\right>$ at the wall are plotted in Fig.\ \ref{powerplots}(b). When the LLJ is above the turbine rotor swept area ($z_\text{jet} > z_h$), there is a significant negative (downward) momentum flux, which extracts the jet's momentum and eliminates it towards the rear of the wind farm. However, when the LLJ is below the turbines ($z_\text{jet} < z_h $), the turbines operate in the negative shear region, and a significant positive entrainment flux is created. As a result, the jet's energy is entrained towards the turbines, and the power production shows an upward trend towards the end of the wind farm, see Fig.\ \ref{powerplots}(a). {{\color{black} In essence, when the LLJ is below the turbine rotor swept area, the momentum deficit by the turbines creates a significant positive turbulent flux from below due to the negative shear.} This enhances the wake recovery further downstream which is further elucidated below.\\
\indent For continuous production of turbulence $\overline{u'w'}\frac{\partial{\overline{u}}}{\partial{z}}$ should be negative. Therefore, in the presence of positive shear ($\frac{\partial{\overline{u}}}{\partial{z}}$ is positive) $\overline{u'w'}$ should be negative to produce turbulence. However, when the shear is negative ($\frac{\partial{\overline{u}}}{\partial{z}}$ is negative) $\overline{u'w'}$ should be positive to sustain turbulence. The tendency of the velocity deficit in the turbine wakes is to create a positive entrainment flux below the hub-height and a negative entrainment flux above hub-height. This leads to the following two scenarios: 
\begin{itemize}
 \item[1.] When the LLJ is above the turbine rotor swept area ($z_\text{jet} > z_h$), LLJ energy is pulled towards the turbines due to the momentum deficit created by the turbines. We have a significant downward entrainment flux, which is utilized by the turbines for power production. Due to which the LLJ strength is reduced. 
 In this case, $\overline{u'w'}$ is negative, and the horizontally averaged $\frac{\partial{\overline{u}}}{\partial{z}}$ is positive, and there is a net negative vertical flux towards the turbines.
 \item[2.] When the LLJ is below the turbine rotor swept area ($z_\text{jet} < z_h$), the high momentum LLJ with the positive entrainment flux from below aid power production. The turbines extract the LLJ energy transported by the positive entrainment fluxes. 
 In this case, $\overline{u'w'}$ is positive and the horizontally averaged $\frac{\partial{\overline{u}}}{\partial{z}}$ is negative. The negative shear created by the wind turbine wakes contributes to the negative shear already present above the LLJ, and this aids power production of downstream turbines.
\end{itemize}

To quantify the turbulence produced by the wakes, the streamwise variation of the horizontally averaged turbulence intensity at the hub-height is plotted in Fig.\ \ref{powerplots}(d). When the LLJ is above the turbine rotor ($z_\text{jet} > z_h$), the turbulence intensity upstream of the farm is 1.97\%, while it is 1.0\% and 0.46\% for $z_\text{jet} \approx z_h$ and $z_\text{jet} < z_h$, respectively. We observe negligible wake turbulence behind the first turbine row in Fig.\ \ref{topview}(d) when the LLJ is below the turbine rotor swept area ($z_h > {z_\text{jet}}$). However, after turbulence is created by the wakes, the turbulence intensity increases to about approximately $4.4\%$ further downstream. It is clear from the above data that there is limited upstream turbulence when the LLJ is below the turbine rotor swept area. Consequently, there is negligible wake recovery until there is a wake generated turbulence. In essence, the negative shear above the jet creates a positive entrainment flux, which increases the turbulence intensity when the LLJ is below the turbine rotor swept area. This accelerates the wake recovery and allows the turbines to extract energy from the jet. The turbulence intensity for the $z_\text{jet} \approx z_h$ case develops in a very similar way as for the ${z_\text{jet}} < z_h$ case as in both cases it is mostly determined by the wake added turbulence. However, when the LLJ is above the turbine rotor swept area, the turbulence intensity inside the wind farm is higher as the atmospheric turbulence interacts with the wind turbine wakes.

\subsection{Energy budget analysis}\label{sec3.3}
To further understand the different processes involved in the power production of a wind farm in the presence of a LLJ we perform an energy budget analysis. The analysis is similar to the budget analysis performed by Allaerts and Meyers \citep{all17} for wind farms in conventionally neutral boundary layers. The steady-state, time-averaged energy equation is obtained by multiplying equation (\ref{eqn2}) with $\widetilde{u}_i$\cite{all17, sag06} and performing time-averaging, which results in:
\begin{equation}
\centering
\begin{split}
 &\overbrace{\overline{u}_j\partial_j\left({\frac{1}{2}\overline{u}_i\overline{u}_i}+\frac{1}{2}\overline{u'_iu'_i}\right)}^\text{Kinetic energy flux}+\overbrace{\partial_j\left(\frac{1}{2}{\overline{u'_ju'_iu'_i}}+\overline{u}_i\overline{u'_iu'_j}\right)}^\text{ Turbulent transport}+\overbrace{\partial_j\left( \overline{u_i\tau_{ij}}\right)}^\text{SGS transport}\\&=\overbrace{-\partial_i{\left(\overline{pu_i}\right)}}^\text{Flow work}+\overbrace{g\beta(\overline{u_i\theta}-\overline{u}_i\theta_0)\delta_{i3}}^{\text{Buoyancy}}+\overbrace{f_c\left(\overline{u}_iU_g\right)\delta_{i2}-f_c\left(\overline{u}_iV_g\right)\delta_{i1}}^{\text{Geostrophic forcing}}\\&+\overbrace{\overline{f_iu_i}}^{\text{Turbine power}}+\overbrace{\overline{\tau_{ij}S_{ij}}}^{\text{Dissipation}}, 
\end{split} \label{eqn8} 
\end{equation}
where the time-averaging is represented by the overline, and $\overline{u'_iu'_j}=\left(\overline{{u_iu_j}}\right) - \overline{{u}_i}~\overline{{u}_j}$ indicates the momentum fluxes. To obtain the total power produced by each row we numerically integrate each term in equation (\ref{eqn8}) around a control volume surrounding each row. The control volume is chosen such that it encloses a row of wind farm, see Fig.\ \ref{cv}. We note here that the fringe layers are not included in the control volume. Performing integration and rearranging equation (\ref{eqn8}) gives,
\begin{equation}
\centering
\begin{split}
 \overbrace{\int^{}_{\forall}\overline{f_iu_i}d\forall}^{\text{$\mathbb{P}$, Turbine power}}&=\overbrace{\int^{}_{\forall}\overline{u}_j\partial_j\left({\frac{1}{2}\overline{u}_i\overline{u}_i}+\frac{1}{2}\overline{u'_iu'_i}\right)d\forall}^\text{$\mathbb{E}_k$, Kinetic energy flux}\\&+\overbrace{\int^{}_{\forall}\partial_j\left(\frac{1}{2}{\overline{u'_ju'_iu'_i}}+\overline{u}_i\overline{u'_iu'_j}\right)d\forall}^\text{$\mathbb{T}_\text{t}$, Turbulent transport}+\overbrace{\int^{}_{\forall}\partial_j\left( \overline{u_i\tau_{ij}}\right)d\forall}^{\text{$\mathbb{T_\text{sgs}}$, SGS transport}}\\&+\overbrace{\int^{}_{\forall}\partial_i{\left(\overline{pu_i}\right)}d\forall}^{\text{$\mathbb{F}$, Flow work}}-\overbrace{\int^{}_{\forall}g\beta(\overline{u_i\theta}-\overline{u}_i\theta_0)\delta_{i3}d\forall}^{\text{$\mathbb{B}$, Buoyancy}}\\&-\overbrace{\int^{}_{\forall}f_c\left(\overline{u}_iU_g\right)\delta_{i2}-f_c\left(\overline{u}_iV_g\right)\delta_{i1}d\forall}^{\text{$\mathbb{G}$, Geostrophic 
 forcing}}-\overbrace{\int^{}_{\forall}\overline{\tau_{ij}S_{ij}}d\forall,}^{\text{$\mathbb{D}$, Dissipation}}
\end{split}\label{eqn9}
\end{equation}

\begin{figure*}[ht!]
	\centering
	\includegraphics[width=0.75\linewidth]{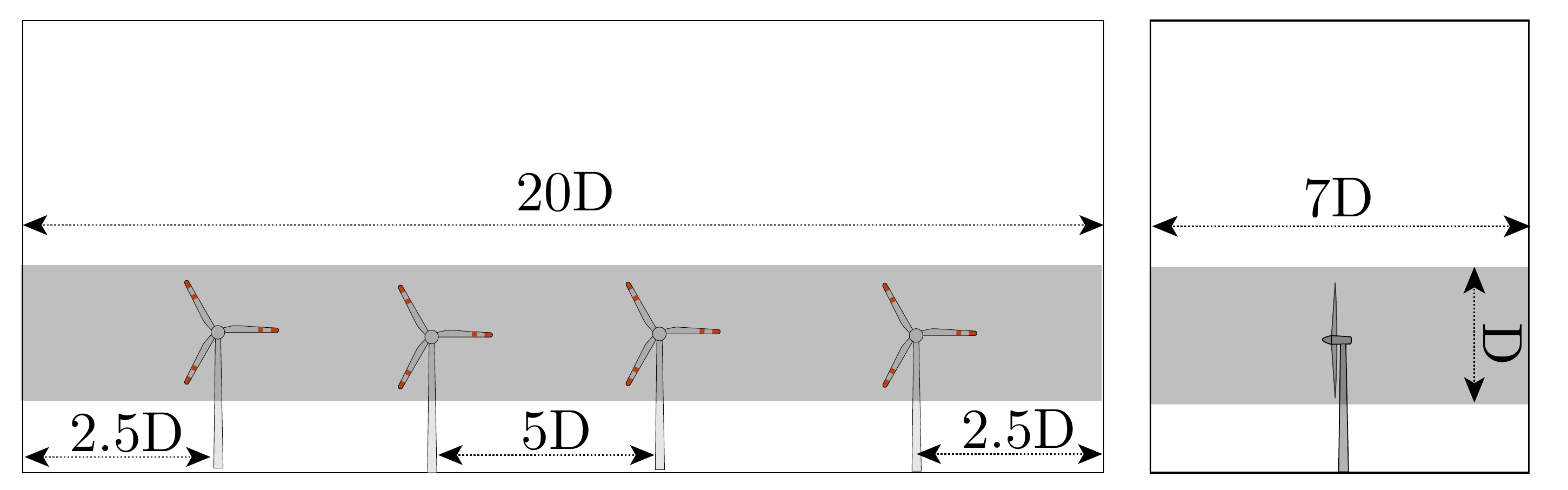}
	\vspace{-0.25cm}
	\caption{The shaded region shows the control volume used in the energy budget analysis. The control volume around each turbine row has dimensions of $7\text{D}\times{20\text{D}}\times{\text{D}}$ in the streamwise, spanwise, and vertical directions, respectively. In the vertical direction the control volume starts at a height of $z_h-D/2$.}
	\label{cv}
\end{figure*}

\begin{figure*}[ht!]
	\centering
	\includegraphics[width=0.85\linewidth]{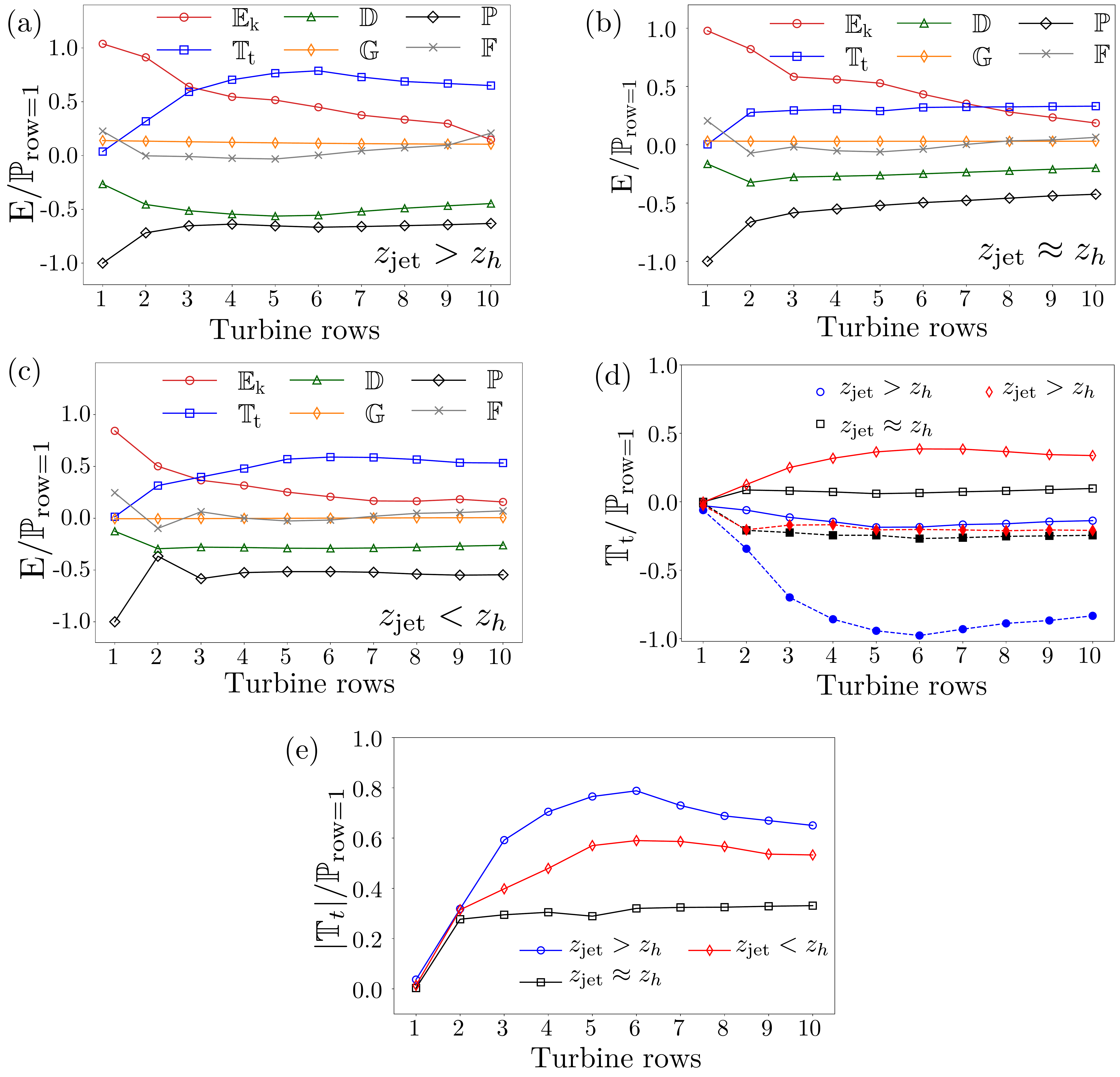}
	\caption{Energy budget for (a) $ z_\text{jet} > z_h$ (b) $ z_\text{jet} \approx z_h$ and (c) $ z_\text{jet} < z_h$. (d) Integrated entrainment flux over top and bottom planes of the control volume. Dashed lines with filled symbols  and solid lines with open symbols represent $\mathbb{T}_t$ on the top and bottom plane, respectively. (e) Normalized net entrainment for different cases.}
	\label{energybudget}
\end{figure*}


\noindent where $\mathbb{P}$ is the power produced by a turbine row, $\mathbb{E}_k$ is the kinetic energy flux containing resolved kinetic energy, $\mathbb{T}_t$ is the turbulent transport, which involves the transport of mean flow energy by turbulence \cite{lum72} and higher-order turbulence terms, $\mathbb{T}_\text{sgs}$ is the mean energy transport by SGS stresses, $\mathbb{F}$ is the flow work, which is the pressure drop across the turbines, $\mathbb{B}$ is the turbulence destruction caused by buoyancy under stable stratification, $\mathbb{G}$ represents the geostrophic forcing driving the flow, and $\mathbb{D}$ is the SGS dissipation.\\
\indent Figure \ref{energybudget}(a), (b), and (c) present the energy budget analysis for cases when the LLJ is above ($z_\text{jet} > z_h$), in the middle ($z_\text{jet}\approx z_h$), or below ($ z_\text{jet} < z_h$) the turbine rotor swept area, respectively. All the terms are normalized by the absolute value of the power produced by the first turbine row. This normalization provides insight into the effect of wake recovery on power production. {\color{black} The SGS transport term $\mathbb{T}_\text{sgs}$ and the buoyancy terms $\mathbb{B}$ are negligible and left out of the plots for brevity.} In both plots, the energy sources are positive, and sinks are negative. Both turbine power and dissipation act as energy sinks in the boundary layer. Fig.\ \ref{energybudget}(a) shows that when the LLJ is above the turbine rotor swept area ($z_\text{jet} > z_h$), the kinetic energy $\mathbb{E}_k$ continuously decreases in the downstream direction. This reduction in the mean kinetic energy is compensated by the turbulent transport term $\mathbb{T}_t$. The turbulent transport slightly reduces after the sixth turbine row due to the reduction in the strength of the LLJ. In this case, the downward entrainment of the fluxes compensates for the decrease in mean kinetic energy. In a fully developed wind farm boundary layer, the power production is completely balanced by the turbulent entrainment from above \cite{cal10, cal10b}. When the turbines operate in the positive shear region, the entrainment from above replenishes the energy extracted by the turbines. In addition to entrainment, the geostrophic forcing $\mathbb{G}$ and pressure drop $\mathbb{F}$ act as an additional energy source. In contrast, turbulence destruction by buoyancy $\mathbb{B}$ and dissipation $\mathbb{D}$ remove energy from the control volume. {\color{black} When the LLJ is above the turbine rotor swept area ($z_\text{jet} > z_h$),} there is positive shear in the boundary layer, due to which there is significant entrainment and wake recovery.\\
When the LLJ is in the middle of the turbine rotor swept area ($z_\text{jet} \approx z_h$), while the kinetic energy $\mathbb{E}_k$ continuously decreases the entrainment $\mathbb{T}_t$ is reduced as the energy in the LLJ is extracted by the upwind turbines reducing the entrainment for the rest of the turbines. Furthermore, turbulent entrainment $\mathbb{T}_t$ is nearly equal to the turbulence dissipation $\mathbb{D}$, this limits the contribution of turbulence to power production.
{\color{black} When the LLJ is below the turbine rotor swept area ($z_\text{jet} < z_h$), the kinetic energy $\mathbb{E}_k$ contribution decreases continuously with downstream position in the wind farm and the turbulent transport $\mathbb{T}_t$ is less than for the case when the LLJ is above the turbines (Fig.\ \ref{energybudget}(c)).} The turbulent transport $\mathbb{T}_t$ is created entirely by the wake turbulence and the momentum deficit created by the turbines. The power production is mainly due to the mean flow energy extraction $\mathbb{E}_k$ and the entrainment due to positive entrainment flux. {\color{black}In essence, the wake recovery and the entrainment due to turbulent transport are affected when the LLJ is below the turbine rotor swept area.}\\
\indent {\color{black} To further elucidate the effect of entrainment on power production, we plot the integrated vertical entrainment flux on the top and bottom planes of the control volume in Fig. \ref{energybudget}(d). Open symbols represent the integrated flux on the bottom plane, and filled symbols represent the integrated flux on the top plane of the control volume. When the LLJ is above the turbine rotor swept area ($z_\text{jet} > z_h$), the negative flux from the top is dominant. However, when the LLJ is below the turbine rotor swept area ($z_\text{jet} < z_h$), there is significant positive flux in the bottom plane indicating positive entrainment from below. This clearly shows that there is significant positive entrainment flux towards the turbine rotors when the LLJ is below the turbine rotor swept area. This is beneficial for the power production of turbines further downstream.}\\
\indent Figure \ref{energybudget}(e) provides a comparison of the net entrainment $|\mathbb{T}_t|$ for all the three cases. {\color{black} The figure shows that entrainment is strongest when the LLJ is above the turbine rotor swept area ($z_\text{jet} > z_h$) and least when $z_\text{jet} \approx z_h$}. {\color{black} When the LLJ is in the middle of the turbine rotor swept area ($z_\text{jet} \approx z_h$) the entrainment is affected as the LLJ energy is mostly extracted by the turbines in the first couple of rows. When the LLJ is below the turbine rotor swept area ($z_\text{jet} < z_h$), there is increased entrainment due to the positive entrainment flux. This creates a stronger turbulent transport $\mathbb{T}_t$ than for the $z_\text{jet} \approx z_h$ case, but not as much as for the case when the LLJ is above the turbine rotor swept area.}

\section{Conclusions}\label{sec4} 
We performed LES of wind farms to study the effect of the LLJ height compared to the turbine-height on the interaction between LLJs and large wind farms, see Fig.\ \ref{fig1}. {\color{black} We considered three scenarios, wherein the LLJ is above, below, and in the middle of the turbine rotor swept area.} We find that the relative power production of the turbines further downstream in the wind farm depends on the jet height relative to the hub-height.} The power production relative to the first-row power is maximum when the LLJ is above the turbine rotor swept area due to higher turbulence intensity below the LLJ, wherein the atmospheric turbulence adds to the turbine wake generated turbulence and leads to a faster wake recovery. {\color{black} However, when the LLJ is below the turbine rotor swept area,} the turbines operate in the negative shear region of the LLJ in which the atmospheric turbulence is limited, and the thermal stability is strong. In the absence of atmospheric turbulence, the wakes are very stable \cite{mao18, kec14}, and wake recovery is slow. However, after the first two turbine rows, the wakes generate sufficient turbulence to promote the wake recovery further downstream. \\
\indent The energy budget analysis reveals that the vertical entrainment dominates the power production when the LLJs are above the turbine rotor swept area. In contrast, when the LLJ is in the middle of the turbine rotor swept area, the jet's energy is extracted by the first turbine row, and the rest of the rows do not directly benefit from the jet. Interestingly, when the LLJ is below the turbine rotor swept area, the mean negative shear and the shear created by the wakes create a positive entrainment flux from below, which helps turbines further downstream to harvest the jet's energy. {\color{black} Although the negative shear above the LLJ creates a positive turbulent entrainment flux, the turbulence production it creates is limited due to the high thermal stratification above the jet, i.e. the flux that is created is smaller than the flux that is created when the LLJ is above the turbines.
}}\\
\indent {\color{black} 
Gutierrez et al.\ \cite{gut17} report the reduction in the turbine loads due to the negative shear in the LLJ and therefore suggest installing turbines such that they are in this region. Our results show that wake recovery is affected when the turbines operate in the negative shear region, and therefore, it might not be beneficial in terms of wake recovery.} Here, we emphasize again that we used a generalized LLJ to study the physical phenomena that result from the interaction of a LLJ with a large wind farm. However, further work will be required to investigate the effect of higher thermal stratification, complex terrain, the strength of the geostrophic wind and its direction, the transition from land to sea on the LLJ characteristics, and how this affects the performance of wind farms.

\section*{author's contributions}
\noindent All authors contributed equally to the manuscript.  

\begin{acknowledgments}
We thank the anonymous referees whose comments have been invaluable in improving the quality of the manuscript. This work is part of the Shell-NWO/FOM-initiative Computational sciences for energy research of Shell and Chemical Sciences, Earth and Live Sciences, Physical Sciences, FOM, and STW. This work was carried out on the national e-infrastructure of SURFsara, a subsidiary of SURF corporation, the collaborative ICT organization for Dutch education and research, and an STW VIDI grant (No. 14868).
\end{acknowledgments}

\section*{Data availability}
\noindent The data that support the findings of this study are available from the corresponding author upon reasonable request.

\section*{References}
\bibliography{literature_windfarms}

\end{document}